\documentclass[journal,comsoc]{IEEEtran}
\usepackage[T1]{fontenc}

\usepackage{cite}[sort]

\usepackage{graphicx}
\usepackage{subcaption}
\ifCLASSINFOpdf
\else
\fi
\usepackage{amsmath}
\usepackage[cmintegrals]{newtxmath}
\begin{document}


\author{Junho~Kweon, Chan-Yong~Jung, Kyung-Bin~Bae,~and~Seong-Ook~Park,~\IEEEmembership{Senior~Member,~IEEE}
	
	\thanks{This work has been submitted to the IEEE for possible publication. Copyright may be transferred without notice, after which this version may no longer be accessible.
		
		The authors are with the Microwave and Antenna Laboratory, School of
		Electrical Engineering, Korea Advanced Institute of Science and Technology
		(KAIST), Daejeon, Republic of Korea (e-mail: 49roinevery@gmail.com; jcy132@kaist.ac.kr;
		carrierbkb@kaist.ac.kr; soparky@kaist.ac.kr)}
}

%
%




\title{Internal Calibration Process Using Chirp Pulses with Application of the Adam Learning Algorithm}

\maketitle

\begin{abstract}
We propose a new internal calibration process using chirp pulses. Our method is utilized to mitigate thermal drift, which is unwanted changes and usually occurs in active elements such as a high power amplifier and low noise amplifier. The proposed method has advantages from two distinct aspects: calibration signal and algorithm. In respect to the calibration signal, our method does not contain an additional signal source because chirp pulses, which are normally used for remote sensing, are used as calibration signals. Moreover, our methods solve the ambiguity problem of analyzing a phase shift which occurs when sinusoidal signals are used as calibration signals. In regards to the algorithm, the Adam learning algorithm avoids learning in the wrong direction, unlike the conventional gradient descent.

Using our method, mathematical forms of received signals are acquired successfully. Our method shows better effectivity compared to the conventional gradient descent algorithm. After compensation, the maximum differences of gain and phase become 0.06 dB and 2.42 degrees, respectively.
\end{abstract}


\IEEEpeerreviewmaketitle

\section{Introduction}
\IEEEPARstart{T}{he} maturity of Synthetic Aperture Radar (SAR) technology has enabled various applications, such as surveillance, ice observation, and topographic analysis \cite{eval,overview}. For precise analysis, the interpretation of SAR data needs the obvious relationship between geophysical parameters and RCS. However, distortion in the system contaminates the SAR data, causing low-quality image and ambiguous estimation of geophysical parameters from SAR imagery. Moreover, as technology is developed, the operating frequency rises, and the size of devices is getting smaller. Therefore, system performance is subject to have unwanted changes. Therefore, internal calibration is needed to compensate for the errors and acquire clear signals.

There are two kinds of internal error of radar systems: static error and dynamic error. A static error may be solved on-ground by measuring and adjusting parameters.
Dynamic error, on the other hand, varies; thus, it needs in-flight calibration \cite{adaptive_antennas,overview,yongchan}. For in-flight calibration, generic internal calibration involves an additional signal source and network \cite{eval,terrasar}. For appropriate calibration, it is crucial to compute the degree of distortion of signals, and the distortion is usually notable on amplitude and phase.

\begin{figure}[h]
	\centering
	\begin{subfigure}{1.0\columnwidth} 
		\includegraphics[width=\columnwidth]{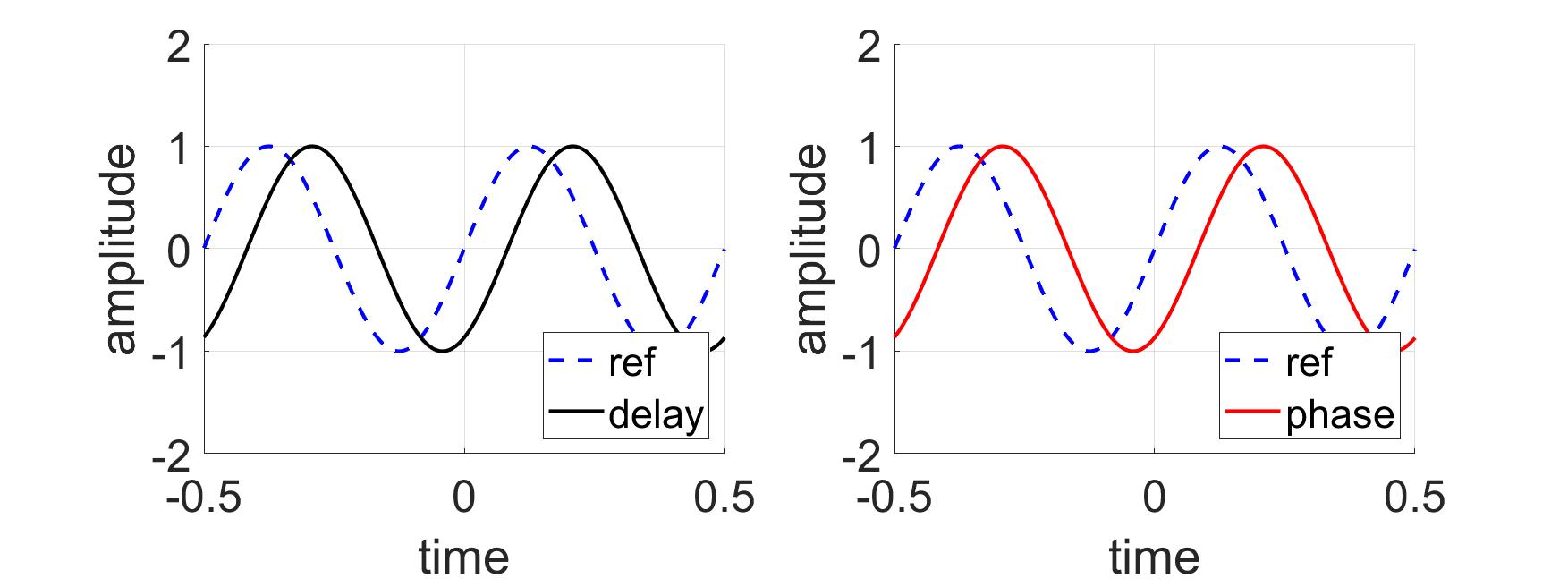}
		\caption{Sinusoidal signals with group delay and phase shift} 
		\label{fig:phaseDifference_sin}
	\end{subfigure}
	\begin{subfigure}{1.0\columnwidth} 
		\includegraphics[width=\columnwidth]{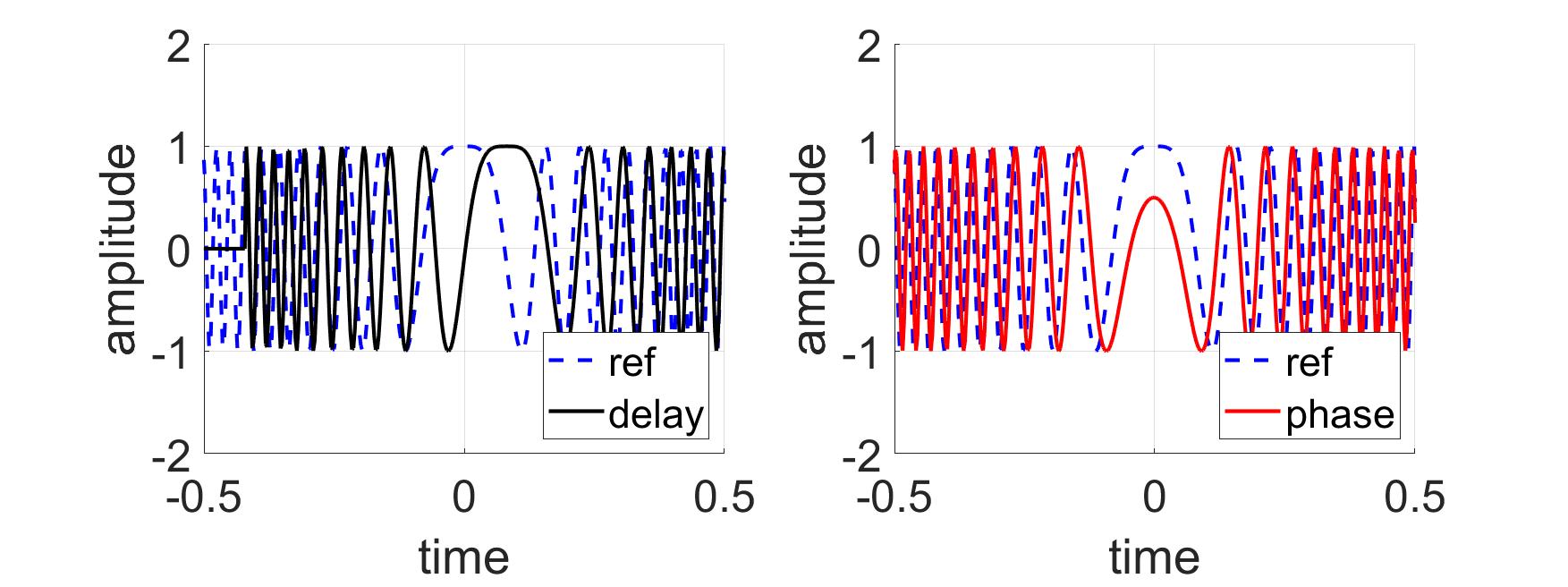}
		\caption{Chirp signals with group delay and phase shift} 
		\label{fig:phaseDifference_chirp}
	\end{subfigure}
	\caption{Difference between group delay and phase shift.} 
	\label{fig:phaseShiftNgroupDelay}
\end{figure}

Previous research used a gradient descent learning algorithm on sinusoidal waves \cite{yongchan}. In that research, to detect the degree of phase and amplitude change, received signals were compared to the reference signals. By doing so, mathematical forms of the sinusoidal waves were derived.  Nevertheless, to apply this idea in the research to real radar systems, there are multiple limitations on two distinct perspectives: signal and algorithm. In regards to the signal, firstly, the CW wave has an ambiguity to choose peak points for the comparison of the phase. Secondly, the pure phase shift between two sinusoidal signals can not be extracted because the phase shift and group delay are outwardly identical. However, in the previous research, the phase shift was calculated by just comparing each closest peak point as if there was no group delay. Thirdly, a sinusoidal wave, which has a singular frequency, cannot represent all frequencies in bandwidth. For the SAR process, radars usually use signals with bandwidth, and predictably gain and phase of a system change depending on the frequency  \cite{software,inta}. Lastly, for a SAR system which uses only chirp signals, an additional signal source may be needed to provide a sinusoidal wave. In respect of the algorithm, the gradient descent algorithm with momentum term was used in previous research \cite{yongchan}. However, this conventional algorithm yields to slow convergence because, when the momentum heads in the wrong direction, it takes more time to come back to the optimization point. 

To solve these problems, we present a new way of internal calibration using chirp signals. This is achievable because a chirp signal, which is used in most SAR systems, shows a difference between group delay and phase shift as plotted in Fig. \ref{fig:phaseShiftNgroupDelay}. Therefore, with the aspects of chirp signals, the phase shift can be distinguished from group delay. At the same time, the Adam learning algorithm yields better effectivity in optimization. 

The remaining sections are organized as follows. In section 2, we discuss the Adam learning algorithm. In section 3, we present how the signals are captured and the internal calibration procedure. In section 4, learning speeds between the Adam learning algorithm and the conventional gradient descent algorithm are compared. In section 5, it is concluded that our method can be preferred by the areas which need short calibration time and compactness.

\section{Adam Learning Algorithm}

\begin{figure}[h]
	\centering
	\includegraphics[width=1.0\columnwidth]{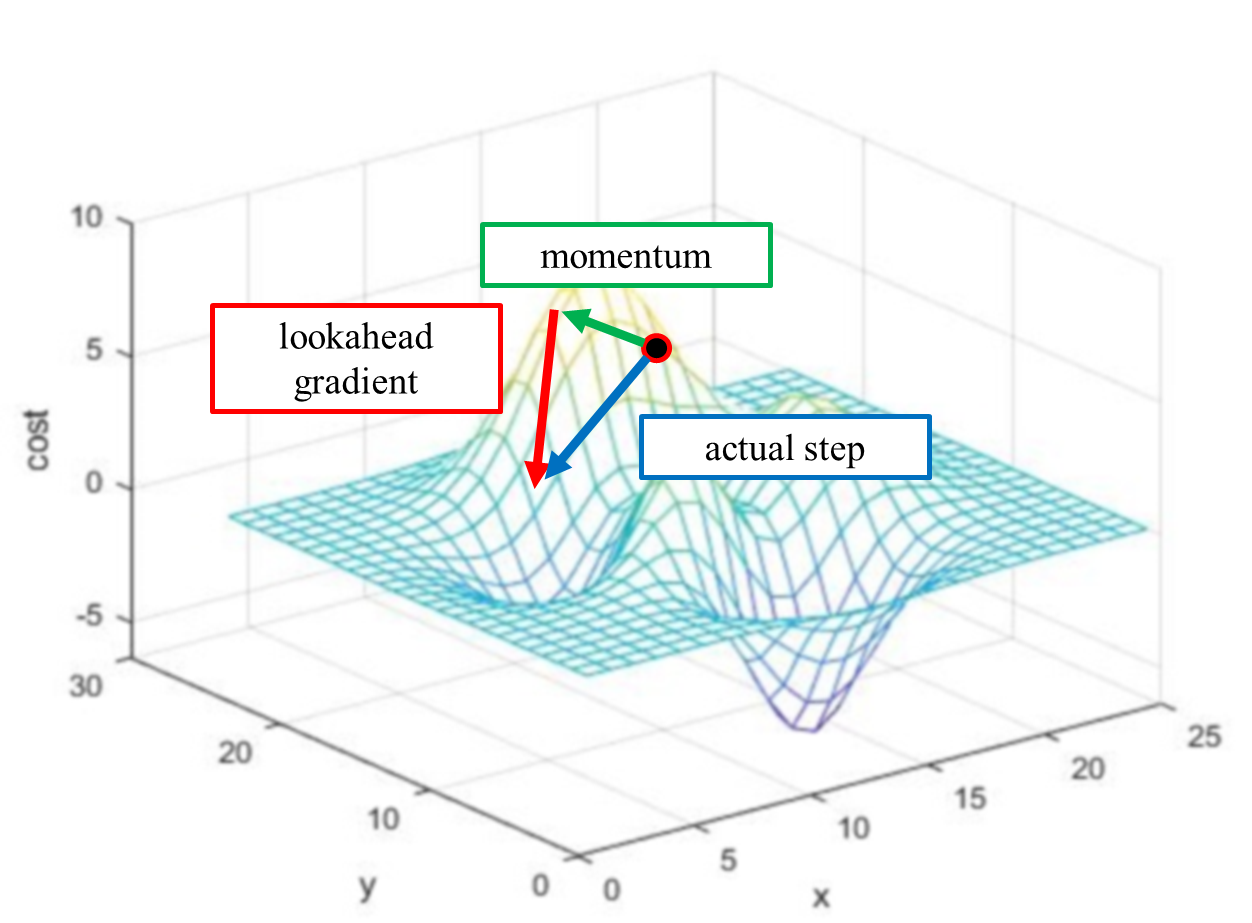}
	\caption{Learning direction of the Adam algorithm \cite{gd}}
	\label{fig:adam}
\end{figure}

To extract a mathematical form of a received signal, amplitude and phase should be found. This can be accomplished by minimizing the error between the mathematical form and the received signal. We apply the Adam learning algorithm, which is a developed version of the conventional gradient descent. The objective of the Adam learning algorithm is to update parameters in a way to minimize the cost function. This algorithm predicts the next two steps: determining the very next location using recent momentum and following the gradient direction from that location. Through this, it avoids detouring to optimize parameters as described in Fig. \ref{fig:adam}. Additionally, this algorithm adjusts the rate of change; the more the parameters have been changed, the more they are optimized. 

In this paper, the amplitude and the phase of the mathematical form of a chirp signal are designated as the parameters to be updated, and the cost function is defined as the error between the mathematical form and a received signal. Eventually, through updates, the mathematical form changes into an imitation of the received signal.

Firstly, the mathematical form of a chirp signal is made using equation \ref{eq:chirp}
\begin{equation}\label{eq:chirp}
	s_l(t)=A\cdot \mathrm{exp}(j\pi K_r t^2+j\pi ft+j\omega)
\end{equation}
where $A, K_r, f,$ and $\omega$ denote amplitude, chirp rate, frequency, and phase, respectively. Initially, $A$ is 1, $K_r$ and $f$ follow the setting of the chirp signal generator, and $\omega$ is set 0. Secondly, the cost function is calculated following the equation 
\begin{equation}\label{eq:costfunction}
	E=\frac{1}{N}\sum_{i=1}^N (d_i-y_i)^2
\end{equation}
where $d_i$ and $y_i$ stand for desired output (received signal) and actual output (mathematical form), and $N$ is the number of samples in a chirp duration. Thirdly, parameters are updated in a way to decrease the cost function. Lastly, when the mathematical form successfully imitates the received signal, amplitude, and phase are acquired.




Equations (\ref{eq:update_exptd_mom_adam})-(\ref{eq:update_theta_adam}) are steps to apply the Adam learning algorithm. Setting for parameters is as $\alpha=1.5e^{-4}$, $\beta_1=0.9$, and $\beta_2=0.999$, where $\alpha$ and $\beta$ denote a stepsize, exponential decay rate for the moment estimates. $m_t$ and $v_t$ are first and second biased moment, and those with hats mean bias-corrected versions. $\nabla_{\theta}J(\theta)$ is the gradient with respect to the cost function.
\begin{gather}
	m_t=\beta_1 m_{t-1}+(1-\beta_1)\nabla_{\theta}J(\theta)
	\label{eq:update_mom_adam}\\
	v_t=\beta_1 v_{t-1}+(1-\beta_2)(\nabla_{\theta}J(\theta))^2
	\label{eq:update_vector_adam}\\
	\hat{m_t}=\frac{m_t}{1-\beta_1^t}\label{eq:update_exptd_mom_adam}\\
	\hat{v_t}=\frac{v_t}{1-\beta_2^t}\label{eq:update_exptd_vector_adam}\\
	\theta=\theta-\frac{\alpha \cdot m_t}{\sqrt{\hat{v_t}+\epsilon}}\label{eq:update_theta_adam}
\end{gather}

\section{Methodology}
\subsection{Signal Acquisition}
\begin{figure}[h]
	\centering
	\includegraphics[width=1.0\columnwidth]{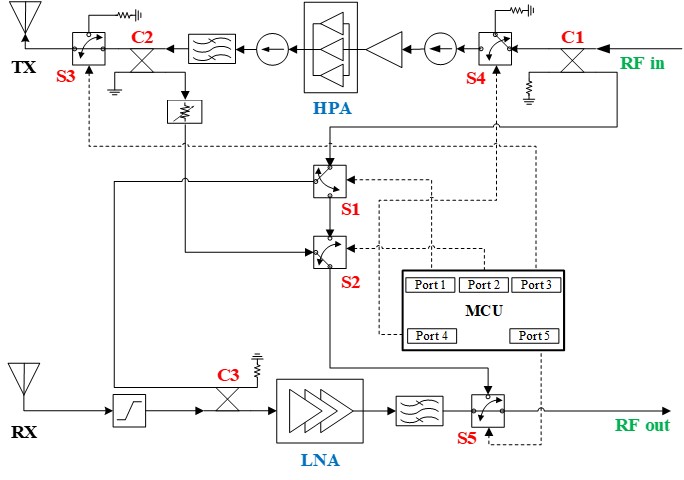}
	\caption{Internal calibration network}
	\label{fig:CalSys}
\end{figure}

For internal calibration, we use a radar system with a calibration network Fig. \ref{fig:CalSys}. In the calibration network, signals can go through different paths by five mechanical switches, which are written as S1-S5. By capturing the signals during switching to compare the signal which went through an amplifier to the one which did not, the gain and phase of the amplifier can be analyzed.

There are four kinds of operational modes for the proposed system. One mode is the normal operation mode where the system transmits and receives signals to acquire target information. The other three modes are internal calibration modes. During normal operation mode, up-converted signals are transmitted after passing HPA and received signals meet LNA before down conversion. During this operation, the system switches its mode between the normal mode and the calibration modes regularly or when internal calibration is needed. 

During the calibration, the system switches its signal path into three distinct paths. In regard to the three paths, path 1 consists of a high power amplifier (HPA) for transmission, path 2 one has a low noise amplifier (LNA), and path 3 does not contain any active amplifiers. The referred signal paths are shown in Table \ref{signalPath}.

\begin{table}[!h]
	\renewcommand{\arraystretch}{1.2}
	\caption{Signal paths of each mode.}
	\label{signalPath}
	\centering
	\begin{tabular}{|c||c|}
		\hline
		Mode	& Path \\
		\hline
		P1		& C1$\rightarrow$S4$\rightarrow$HPA$\rightarrow$C2$\rightarrow$S2$\rightarrow$S5		\\
		\hline
		P2		& C1$\rightarrow$S1$\rightarrow$C3$\rightarrow$LNA$\rightarrow$S5			\\
		\hline
		P3		& C1$\rightarrow$S1$\rightarrow$S2$\rightarrow$S5				\\
		\hline	\hline
		Transmit& C1$\rightarrow$S4$\rightarrow$HPA$\rightarrow$C2$\rightarrow$S3$\rightarrow$TX \\ 
		\hline
		Receive	& RX$\rightarrow$C3$\rightarrow$LNA$\rightarrow$S5	\\
		\hline
	\end{tabular}
\end{table}

To acquire signals, an internal calibration system is used to make different signal paths. On the received signals, the Adam learning algorithm is applied to extract mathematical forms in order to calculate the amplitude and phase of HPA and LNA. This is repeated when the temperature of those active elements is changed by 0.2$^\circ$C. Comparing to the reference gain and phase of the HPA and LNA, their thermal drift can be compensated.

\subsection{Internal Calibration Procedure}

\begin{figure}[h]
	\centering
	\includegraphics[width=1.0\columnwidth]{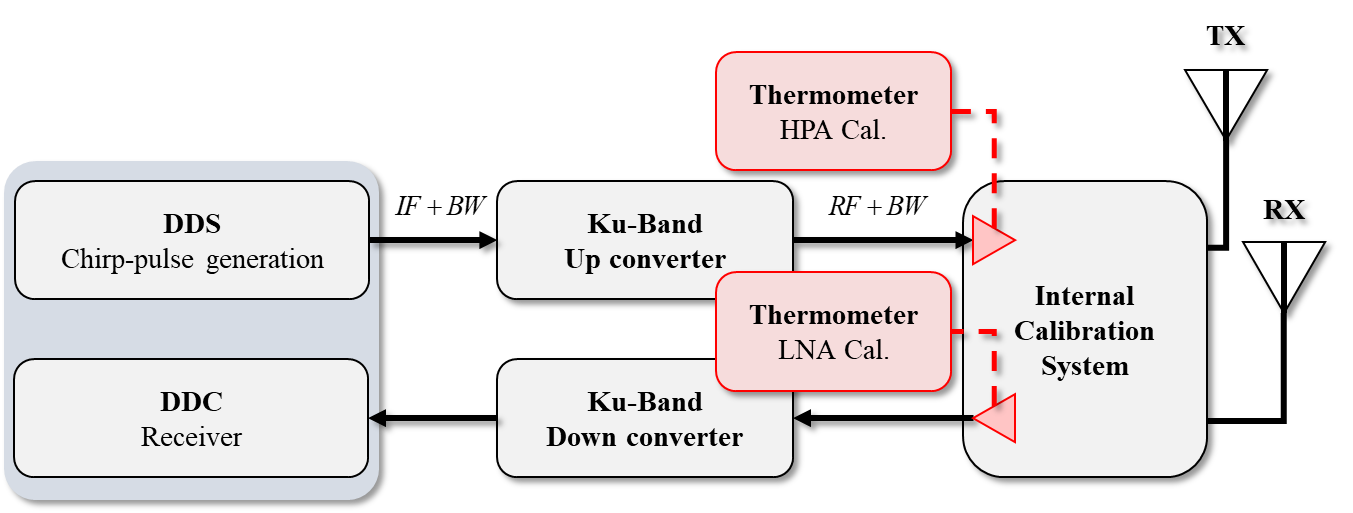}
	\caption{Measurement setting for internal calibration}
	\label{fig:measSetting}
\end{figure}

Using the Adam learning algorithm, the amplitude and phase of the received signal are acquired. By comparing parameters from the signal which goes through an amplifier and the other which does not, the gain and phase of an amplifier are calculated. 
The gain and phase of the system can change because of thermal drift in active elements such as HPA and LNA. Thus, signals were acquired repeatedly by measuring the temperature of the HPA and LNA simultaneously. Fig. \ref{fig:measSetting} shows our setting for the signal acquisition and the temperature measurement, and a Ku-band system is used for evaluation.

By comparing gain and phase to their reference values, calibration factors are derived following equations (\ref{eq:gainComp}) and (\ref{eq:phaseComp}). On those equations, subscript $H$ and $L$ indicate HPA and LNA, respectively. Parameters with the superscript $r$ are reference values. $k$ and $\theta$ are calibration factors for compensation. With the calibration factors, the unwanted changes in gain and phase of the active elements are compensated.

\begin{subequations}\label{eq:gainComp}
	\begin{equation}
		G_H^r=G_H k_H
	\end{equation}
	\begin{equation}
		G_L^r=G_L k_L
	\end{equation}
\end{subequations}

\begin{subequations}\label{eq:phaseComp}
	\begin{equation}
		\phi_H^r=\phi_H + \theta_H
	\end{equation}
	\begin{equation}
		\phi_L^r=\phi_L + \theta_L
	\end{equation}
\end{subequations}

\section{Results}
Chirp pulses, which are generated following Table \ref{chirpSetting}, go through three paths P1, P2, and P3. The signals are received during switching so that signals from different paths are observable at the same time. The evaluation of the learning algorithm is carried out with the consideration of hardware offset factors like in the previous paper \cite{yongchan}.

\begin{table}[!h]
	\renewcommand{\arraystretch}{1.2}
	\caption{Chirp parameter setting.}
	\label{chirpSetting}
	\centering
	\begin{tabular}{|c|c||c|c|}
		\hline
		\bf{Parameter}				& \bf{Value}& \bf{Parameter} 	& \bf{Value}	 	\\
		\hline
		Pulse Repetition Interval	& 20 $\mu$s &	Bandwidth 		& 80 MHz 				\\
		\hline
		Sampling frequency			& 350 MHz	&	Pulse duration 	& 1.001 $\mu$s 	\\
		\hline
	\end{tabular}
\end{table}

\begin{figure}[h]
	\centering
	\begin{subfigure}{0.49\columnwidth} 
		\includegraphics[width=\columnwidth]{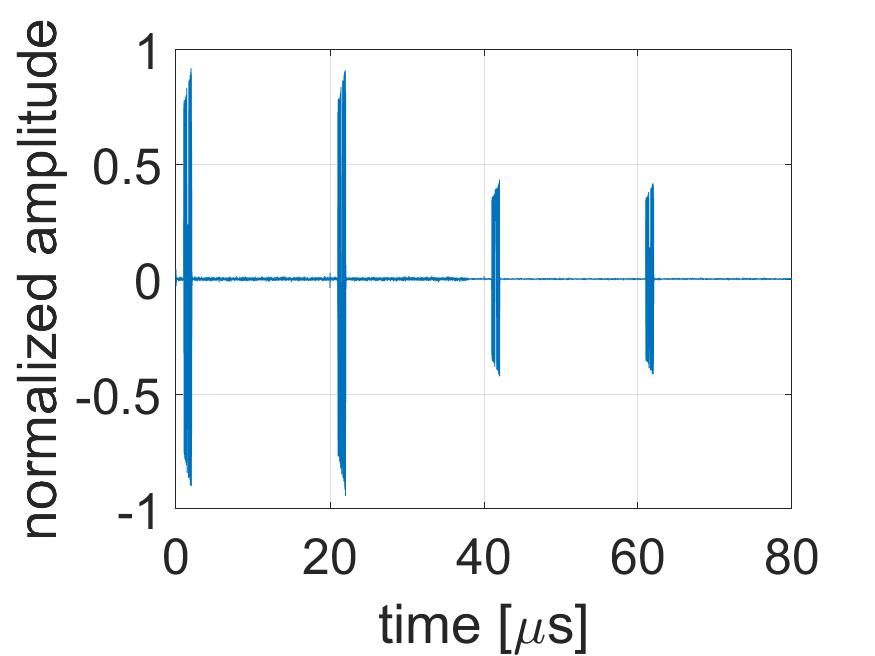}
		\caption{switching from P2 to P3} 
		\label{fig:swP2toP3}
	\end{subfigure}
	\begin{subfigure}{0.49\columnwidth} 
		\includegraphics[width=\columnwidth]{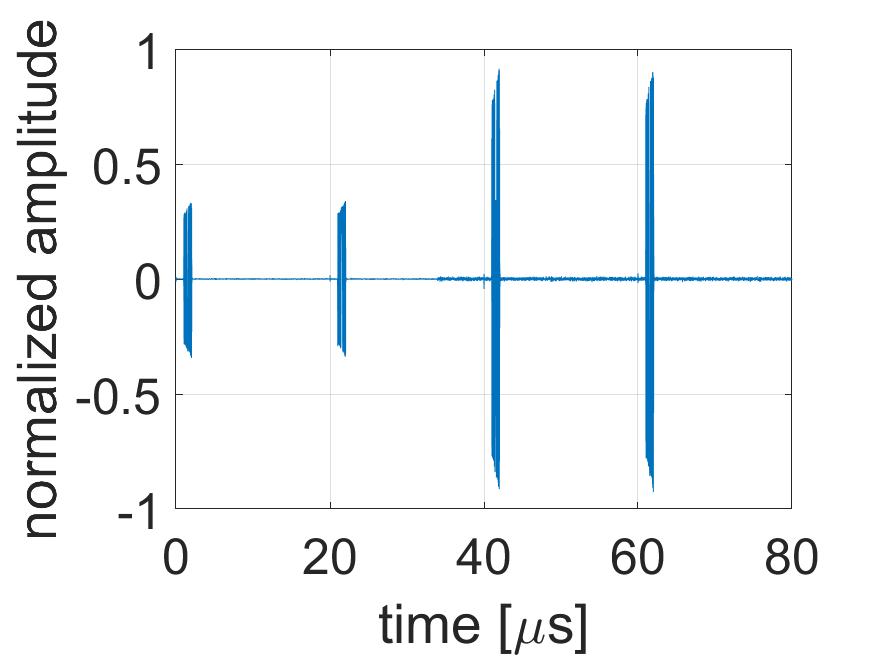}
		\caption{switching from P3 to P1} 
		\label{fig:swP3toP1}
	\end{subfigure}
	\caption{Received chirp pulses during switching.}
	\label{fig:sw}
\end{figure}

Fig. \ref{fig:sw} shows the received chirp pulses during switching. By comparing chirps from different paths, gain and phase of HPA and LNA are evaluated. 

\begin{figure}[h]
	\centering
	\includegraphics[width=0.6\columnwidth]{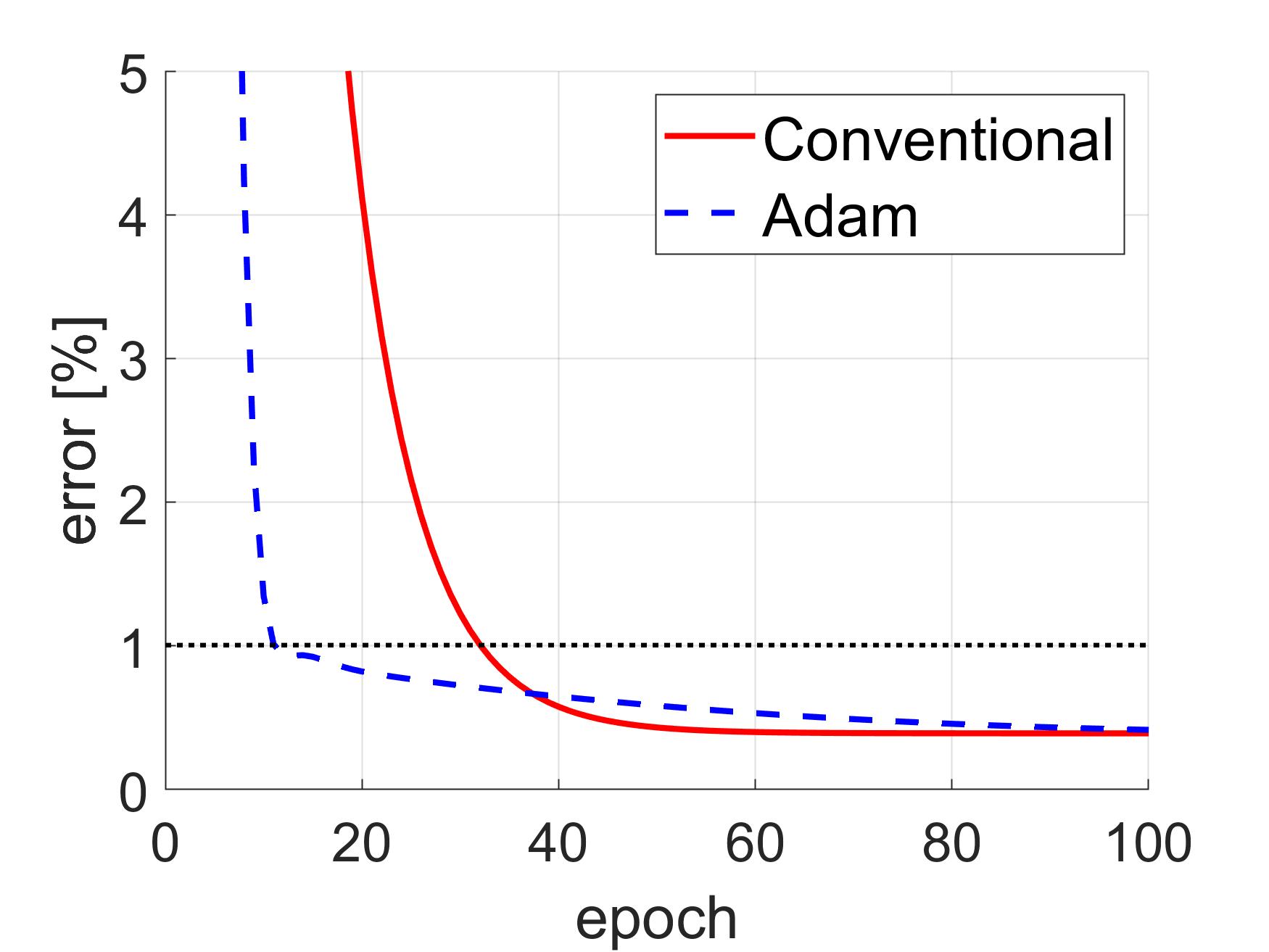}
	\caption{Learning speed comparison}
	\label{fig:compare_learning}
\end{figure}

As shown in Fig. \ref{fig:compare_learning}, both of the algorithms show convergence with small change across epochs. In the figure, the solid line is the conventional gradient descent algorithm, and the dashed line is the Adam learning algorithm. Learning speed is designated as the epoch when error becomes 1 \% of the initial error value. Therefore, in regard to the learning speed, the Adam learning algorithm shows faster learning than the conventional gradient descent algorithm.

\begin{figure}[h]
	\centering
	\begin{subfigure}{0.49\columnwidth} 
		\includegraphics[width=\columnwidth]{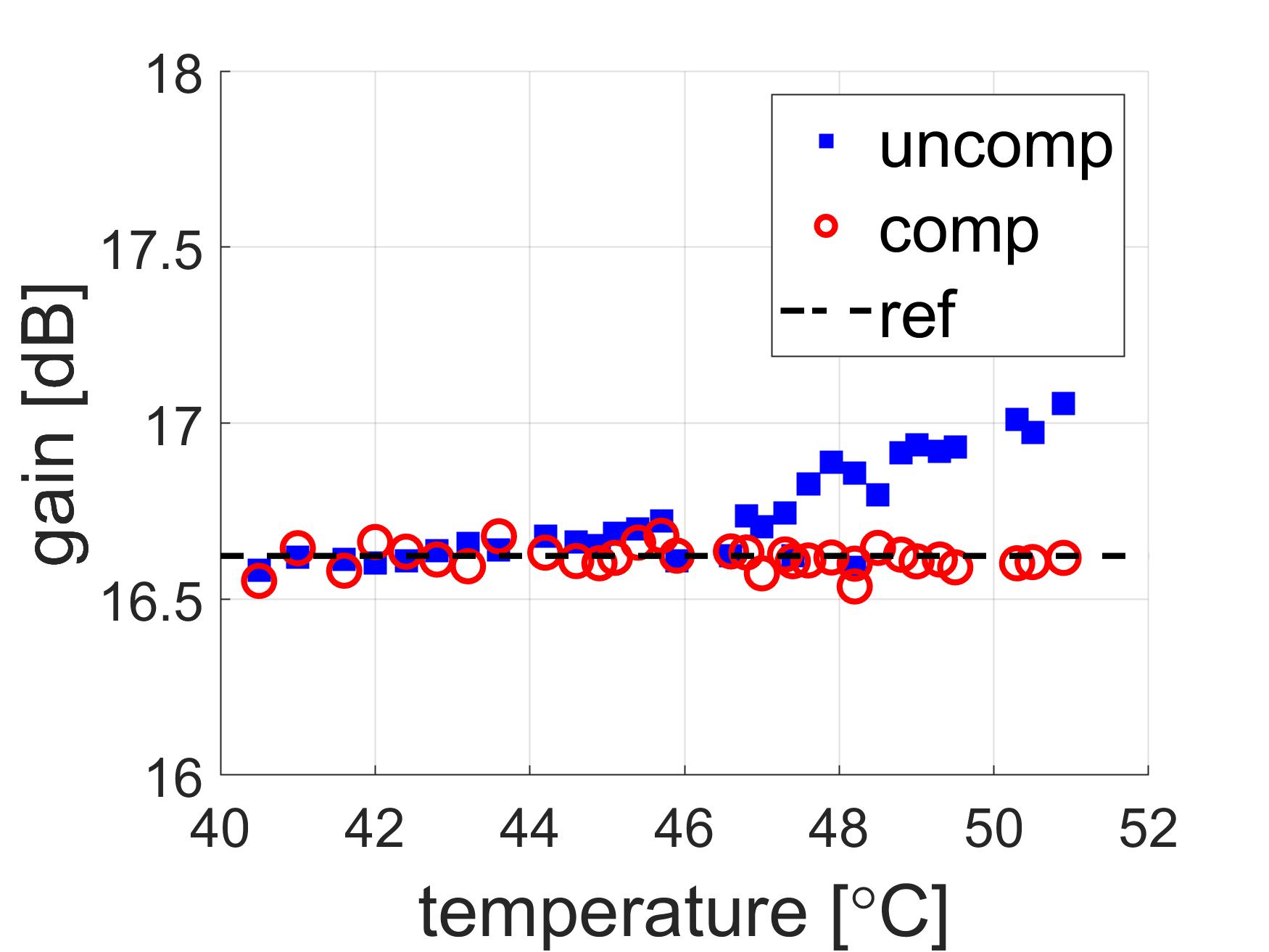}
		\caption{Gain of HPA} 
		\label{fig:HPA_gain}
	\end{subfigure}
	\begin{subfigure}{0.49\columnwidth} 
		\includegraphics[width=\columnwidth]{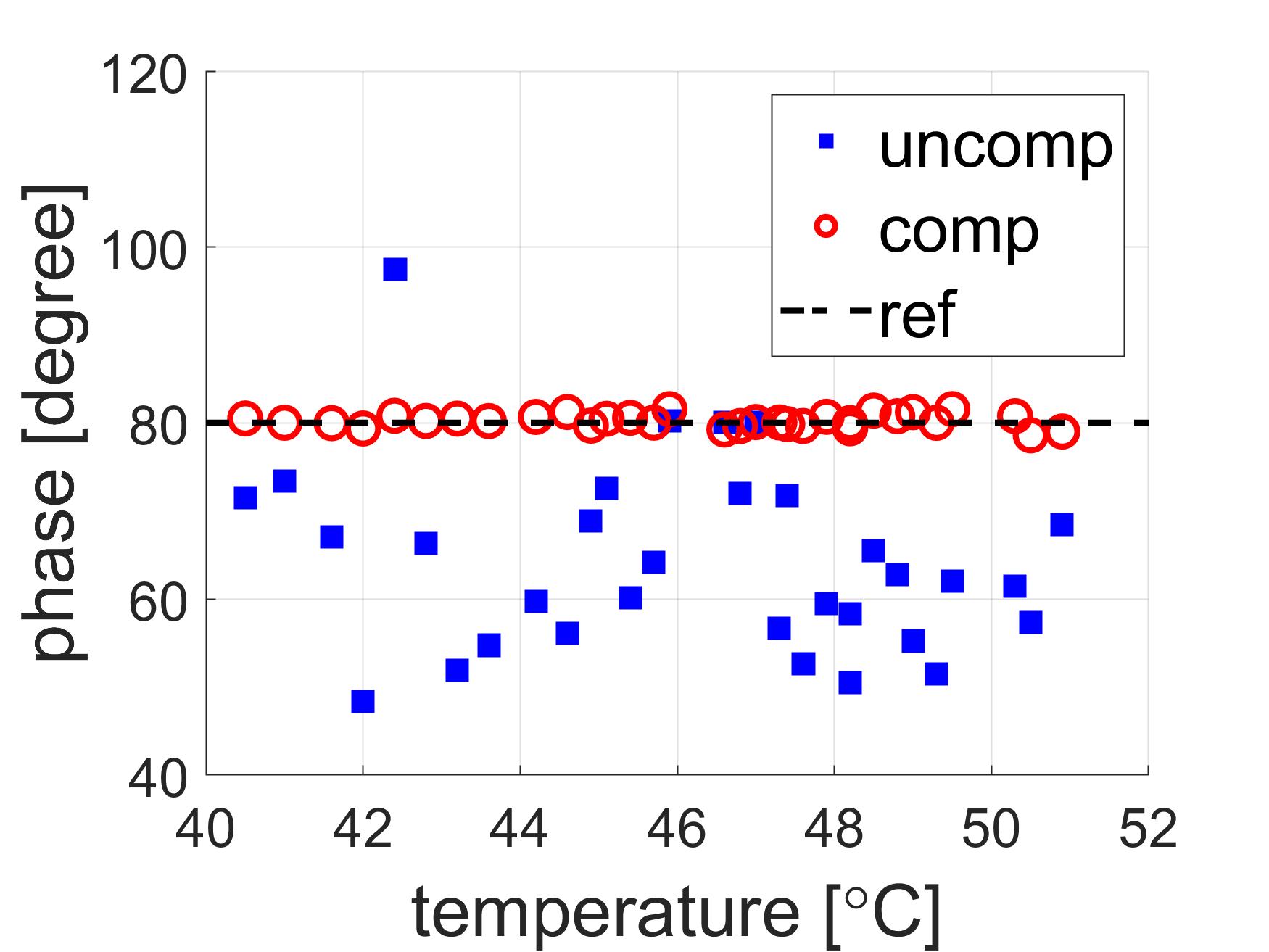}
		\caption{Phase of HPA} 
		\label{fig:HPA_phase}
	\end{subfigure}
	\begin{subfigure}{0.49\columnwidth} 
		\includegraphics[width=\columnwidth]{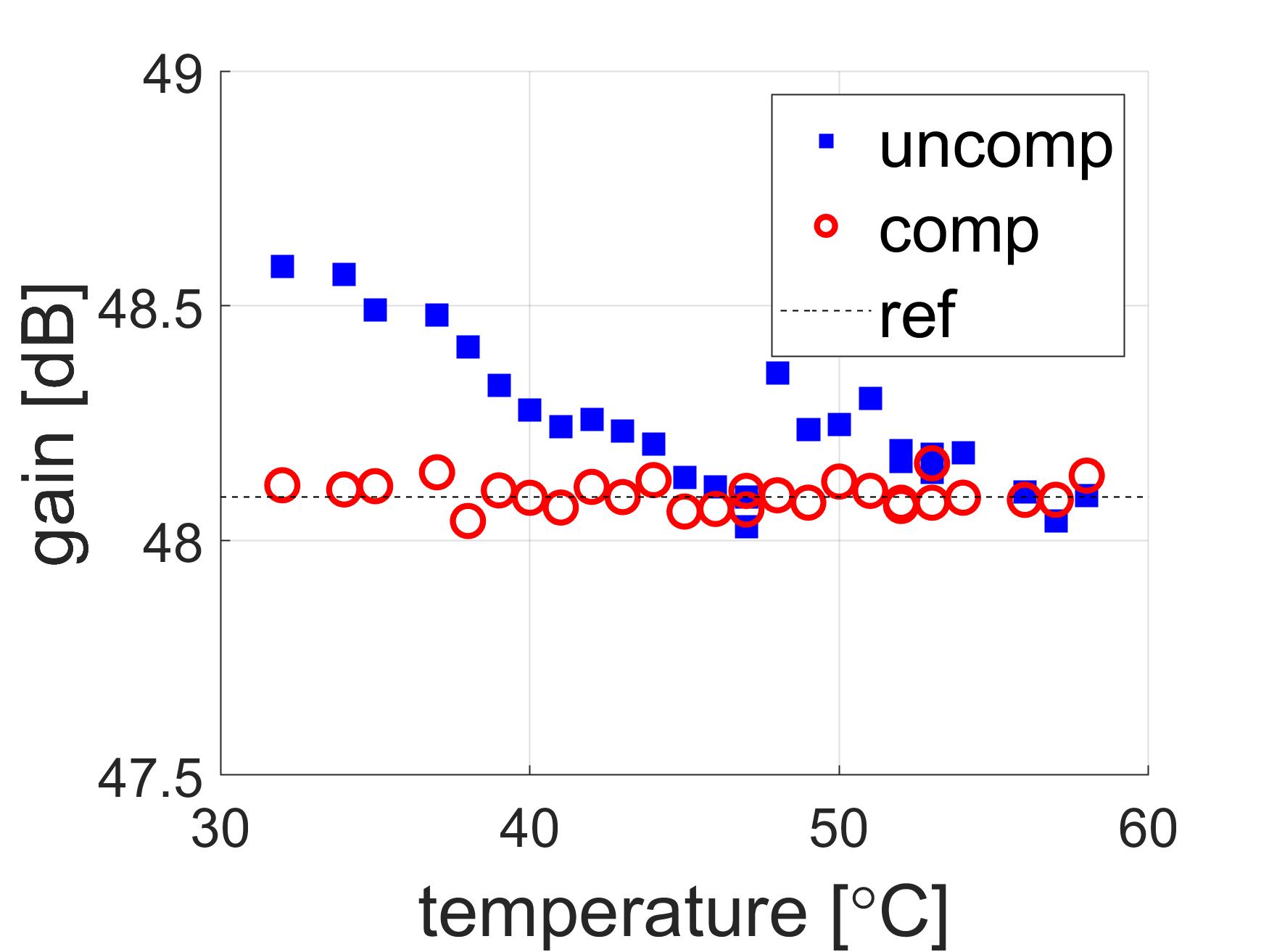}
		\caption{Gain of LNA} 
		\label{fig:LNA_gain}
	\end{subfigure}
	\begin{subfigure}{0.49\columnwidth} 
		\includegraphics[width=\columnwidth]{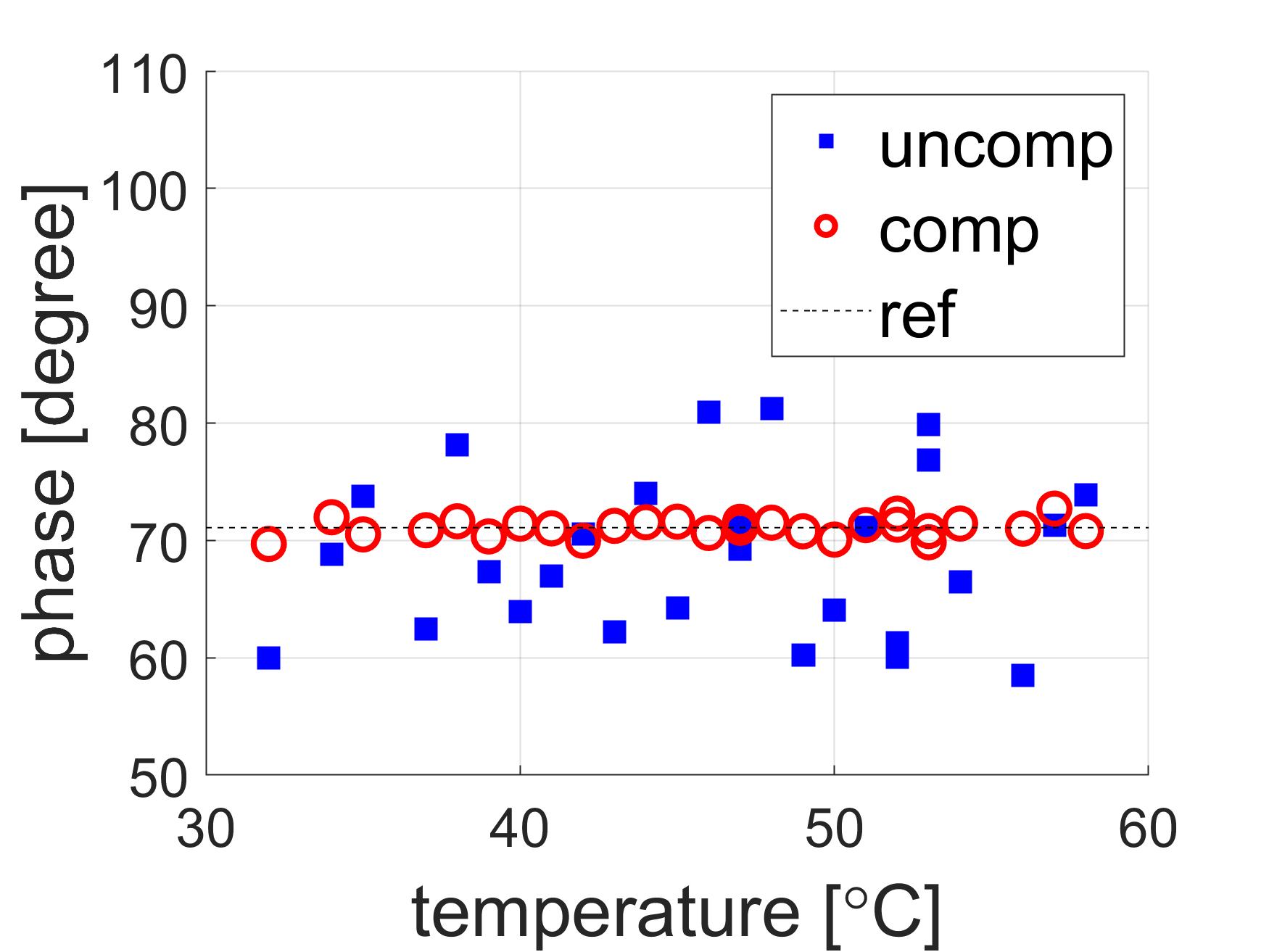}
		\caption{Phase of LNA} 
		\label{fig:LNA_phase}
	\end{subfigure}
	\caption{Measured and compensated values across temperatures.}
	\label{fig:gainNphase}
\end{figure}

In Fig. \ref{fig:gainNphase}, measured values of gain and phase, which are plotted as filled squares, vary across temperatures due to thermal drift. After our proposed internal calibration process, compensated gain and phase (plotted as red circles) are aligned with the reference value. This result validates that our method solves the distortion problem which is caused by the thermal drift.

Numerical results of internal calibration are shown in TABLE \ref{maxDiff_gainNphase}. Obviously, compensation using our method decreased the differences between practical (measured) and reference values. These results validate our method comparing to other calibration results: result A is from \cite{individual} and B is from \cite{final_results}, respectively.

\begin{table}[!h]
	\renewcommand{\arraystretch}{1.2}
	\caption{Max differences between measured and reference values.}
	\label{maxDiff_gainNphase}
	\centering
	\begin{tabular}{|c||c|c||c|c|}
		\hline
		\bf{Element}& \multicolumn{2}{|c||}{\bf{HPA}} & \multicolumn{2}{|c|}{\bf{LNA}} \\
		\hline
		\bf{Parameter}	& \bf{Gain [dB]} & \bf{Phase [deg]} 	& \bf{Gain [dB]}	& \bf{Phase [deg]} \\
		\hline
		\bf{Uncomp}		& 0.43 		& 31.73			& 0.49		& 12.60 \\
		\hline
		\bf{Comp}		& 0.06 		& 2.02			& 0.05		& 2.42 	\\
		\hline \hline
		\bf{Result A}
		& 0.03 		& 0.6			& 0.06		& 0.5 	\\
		\hline
		\bf{Result B}
		& 0.2 		& 2.0			& 0.2		& 2.0 	\\
		\hline
	\end{tabular}
\end{table}

\section{Conclusion}
Though the compensated values of gain and phase are slightly different from reference values, our method improved the stability of the radar system fairly well compared to other research. 
However, there is a limitation in our method that this calibration yields non-perfect compensation if path 3 has distortion. This is caused by the fact that signals from path 3, if not ideal, are used as reference signals. There may be an unavoidable noise issue in path 3, but coherent summation can mitigate the noise effect concerning white Gaussian noise. 

Most radar systems cannot perform remote sensing during the calibration mode. However, our proposed algorithm allows this blind-time to become shorter than the conventional gradient descent algorithm. It is a significant difference because operating time should be guaranteed. The fast speed of the proposed algorithm will improve online calibration for radar systems, such as military and satellite surveillance, which require short-time calibration. Furthermore, our algorithm is not frequency-dependent, so it can be applied to the next frequency-band technologies.



\ifCLASSOPTIONcaptionsoff
  \newpage
\fi

\end{document}